\documentclass{aastex}
\usepackage{emulateapj5}

\newcommand{\g}{\gamma}

\newcommand{\e}{\epsilon}
\newcommand{\ep}{\epsilon^\prime}

\newcommand{\fes}{f^s_{\epsilon_s}}
\newcommand{\fet}{f^{EC}_{\epsilon_{EC}}}
\newcommand{\fessc}{f^{SSC}_{\epsilon_{C}}}

\shorttitle{Radiation Processes in X-ray Jets}
\shortauthors{Dermer \& Atoyan}

\received{}
\begin{document}

\title{Nonthermal Radiation Processes in X-ray Jets}

\author{Charles D. Dermer\altaffilmark{1} \& Armen Atoyan\altaffilmark{2}}
%\affil{E. O. Hulburt Center for Space Research, Code 7653,\\
%Naval Research Laboratory, Washington, DC 20375-5352}
\altaffiltext{1}{E. O. Hulburt Center for Space Research, Code 7653,
Naval Research Laboratory, Washington, DC 20375-5352}
\altaffiltext{2}{Centre de Recherches Math\'ematiques, 
Universit\'e de Montr\'eal, Montr\'eal, Canada H3C 3J7}
%\email{dermer@gamma.nrl.navy.mil}

%\author{Armen Atoyan}
%\affil{Centre de Recherches Math\'ematiques, Universit\'e de Montr\'eal 
%\\Montr\'eal, Canada H3C 3J7}

%\email{atoyan@crm.umontreal.ca}

\begin{abstract}
Analytic approximations for synchrotron, synchrotron self-Compton
(SSC), and external Compton (EC) processes are used to constrain model
parameters for knot and hot-spot emission in extended jets of radio
galaxies.  Equipartition formulas are derived that relate the
Doppler factor $\delta$ and comoving magnetic field $B$ assuming a
nonthermal synchrotron origin of the radio emission, and synchrotron,
SSC and EC origins of the X-ray emission.  Expressions are also
derived for $\delta$ and $B$ that minimize the total jet powers of the
emitting region in synchrotron, SSC and EC models for the X-ray
emission. The results are applied to knot WK7.8 of PKS 0637-752.
Predictions to test two-component synchrotron and EC models are made
for {\it Chandra} and {\it GLAST}.
\end{abstract}


\section{Introduction }

The ability of the {\it Chandra X-ray Observatory} to resolve knots
and hot spots in radio jets
%\footnote{A table of jets with extended
%X-ray emission, compiled by D.\ Harris, is given at
%http://hea-www.harvard.edu/XJET/} 
has opened a new chapter in jet research \citep{sta04}.
The radio emission in the extended jets on multi-kpc -- Mpc size
scales is almost certainly nonthermal synchrotron radiation, but the
origin of the X-ray emission is controversial. In many knots and hot
spots, the spectral energy distribution (SED) at X-ray energies is a
smooth extension of the radio and optical fluxes, so that a
synchrotron origin of the X-ray emission is implied. In other cases,
the X-ray flux exceeds the level implied by smoothly extending the
radio/optical SED. But even in these cases, a one-component
synchrotron interpretation may be possible
\citep{da02}, and a two-component synchrotron model can be preferred
on energetic and spectral grounds \citep{ad04}.

Besides the synchrotron mechanism, two other nonthermal processes are
often considered to account for the X-ray fluxes observed from the
knots and hot spots of radio jets, namely the synchrotron self-Compton
(SSC) and the external Compton (EC) processes \citep{hk02}. The target
photons for the EC model can be CMBR photons \citep{tav00}, or nuclear
jet radiation \citep{bru01}.  The EC model involving CMBR target
photons is the currently favored interpretation for quasar X-ray knots
and hot spots where the X-ray spectrum is not a smooth extension of
the radio/optical spectrum \citep{cgc01,sam04}. In this model, the
X-ray emission from knots such as WK7.8 of PKS 0637-752 is argued to
be due to CMB photons that are Compton-upscattered by nonthermal
electrons from kpc-scale emitting regions in bulk relativistic motion
at distances up to several hundred kpc from the central engine.

In a recent paper \citep{ad04}, we have addressed difficulties of the
X-ray EC model to explain the opposite behaviors of the X-ray and
radio spatial profiles. This model requires large energies,
particularly in debeamed cases where the observer is outside the
Doppler beaming cone.  Here we concentrate on radiation from
the extended X-ray jets, and provide equations suitable for observers
to interpret multiwavelength X-ray data of knot and hot-spot emission
with synchrotron, EC and SSC models,  and evaluate jet
powers. Application to knot WK7.8 of PKS 0637-752 is used to
illustrate the results. 

\section{Approximate Expressions for Radiation Processes}

We consider a spherical blob with radius $r_b$ and comoving volume
$V_b^\prime = 4\pi r_b^3/3$ that moves with bulk Lorentz factor
$\Gamma= (1-\beta^2)^{-1/2}$. Emission from the blob is observed at
angle $\theta_{jet} = \arccos \mu$ with respect to the jet
direction. The Doppler factor $\delta =[\Gamma(1-\beta\mu)]^{-1}$.
A randomly oriented magnetic field with intensity $B$ is assumed to
fill the volume of the blob. Nonthermal relativistic electrons are
assumed to be uniformly distributed throughout the blob with an
isotropic pitch-angle distribution and a Lorentz-factor distribution
$N^\prime_e(\gamma)$, where $N^\prime_e(\gamma) d\gamma$ is the number
of electrons with comoving Lorentz factors $\gamma$ between $\gamma$
and $\gamma + d\gamma$.

For power-law electrons in the range $\gamma_1 \leq \gamma
\leq \gamma_2$, 
\begin{equation}
N^\prime_e(\gamma) = K\gamma^{-p}H(\gamma;\gamma_1,\gamma_2)\;,
\label{Nprime}
\end{equation} 
where the Heaviside function $H(\gamma;\gamma_1,\gamma_2) = 1$ if
$\gamma_1
\leq \gamma \leq \gamma_2$, and $H(\gamma;\gamma_1,\gamma_2) = 0$ otherwise.
Normalizing to the total comoving electron energy $W_e^\prime
= m_ec^2 \int_1^\infty d\gamma \gamma N^\prime_e(\gamma)$ implies 
$K \cong (p-2)W_e^\prime / (m_ec^2 \g_1^{2-p})$ when $p > 2$ and $\g_2 \gg \g_1$.
%\begin{equation}
%K \cong {(p-2)W_e^\prime \over m_ec^2 \g_1^{2-p}}\;.
%\label{K}
%\end{equation}

The $f_\e = \nu F_\nu$ synchrotron radiation spectrum is approximated by
the expression
\begin{equation}
f_\e^{s} \cong {\delta^4\over 6\pi d_L^2} \; c\sigma_{\rm T} u_B 
\gamma_s^3 N_e^\prime (\gamma_s)\;,\;\gamma_s =  
\sqrt{{\e_z\over \delta\e_B}}\;,
\label{fes}
\end{equation}
where $\e = h\nu/m_ec^2$, $\e_z = (1+z)\e$, $u_B = B^2/8\pi$
is the magnetic field energy density, $d_L$ is the luminosity distance
of the source at redshift $z$, and $\e_B = B/B_{cr}$, where the
critical magnetic field $B_{cr} = m_e^2 c^3/e\hbar = 4.414\times
10^{13}$ G.

The expression
\begin{equation}
f_\e^{EC} \cong {\delta^6\over 6\pi d_L^2}\; c\sigma_{\rm T} u_* 
\gamma_{EC}^3 N_e^\prime (\gamma_{EC})\;,\; 
\gamma_{EC} =  {1\over \delta} 
\sqrt{{\e_z\over 2\e_*}}\;
\label{fet}
\end{equation}
is used to approximate the $\nu F_\nu$ spectrum in the EC process
\citep{der95,dss97}. Here we assume that in the stationary frame, 
the radiation field is isotropic and monochromatic with dimensionless
photon energy $\e_*$ and energy density $u_*$ of the target photon
field. Moreover, all scattering is assumed to take place in the
Thomson regime, which holds for X-ray emission from Compton-scattered
CMBR (\citet{gkm01} treat scattering in the Klein-Nishina regime). For
the CMBR, we let $\e_* = 2.70 k_{\rm B}T_{CMB}(1+z) /m_ec^2 =
1.24\times10^{-9}(1+z)$, and $u_* = 4\times 10^{-13}(1+z)^4$ ergs
cm$^{-3}$.

For the SSC spectrum, the internal photon target density is the
synchrotron radiation spectrum $ n_{ph}(\e) \cong r_b\dot N_s(\e)/ c
V^\prime_b $, where $\dot N_s(\e)$ is the synchrotron spectral
emissivity. Using a $\delta$-function approximation for the
Compton-scattered spectrum in the Thomson regime gives
\begin{equation}
f_\e^{SSC} \cong {\delta^4\over 9\pi d_L^2} {c\sigma_{\rm T}^2 
r_b u_B K^2\over V_b^\prime} \gamma_s^{3-p} \Sigma_{\rm C}\;
\label{fessc}
\end{equation}
\citep{dss97},
where the Compton-synchrotron logarithm 
$\Sigma_{\rm C} = \ln (a_{max}/a_{min})$,  
$a_{max}= \min(\e_B\gamma_2^2,\ep/\gamma_1^2,\e^{\prime -1})$
and 
$a_{min}=\max(\e_B\gamma_1^2,\ep/\gamma_2^2 )$
%\ep/\gamma_1^2,\e^{\prime -1})\over
%\max(\e_B\gamma_1^2,\ep/\gamma_2^2 )}\right ]
%\begin{equation}
%\Sigma_{\rm C} = \ln \left [ {\min(\e_B\gamma_2^2,
%\ep/\gamma_1^2,\e^{\prime -1})\over
%\max(\e_B\gamma_1^2,\ep/\gamma_2^2 )}\right ]
%\label{sigmac}
%\end{equation}
\citep{gou79}, and $\ep = \e_z/\delta$.

Eqs.\ (\ref{fes}), (\ref{fet}), and (\ref{fessc}) are used to
approximate the synchrotron, EC, and SSC $\nu F_\nu$ fluxes,
respectively, and are accurate to better than 50\% when $2\lesssim p
\lesssim 3.5$ in the power-law portion of the spectrum, compared to
more precise treatments \citep{bg70,gou79}.

\section{Constraints on $\delta$ and B}

The total particle energy density $u_{par}$ is related to $u_B$
through the relation $u_{par} =W_{par}^\prime/V_b^\prime = k_{eq}
u_B$, where $W^\prime_{par}=W^\prime_e(1+k_{pe})$ is the total
comoving particle energy, $k_{eq}\equiv (1+k_{pe})\tilde k_{eq}$ is a
parameter that measures the deviation of total particle to
magnetic-field energy density, and $k_{pe}$ is the ratio of total
proton energy to total electron energy. The term $\tilde k_{eq}=
W_e^\prime/ (V_b^\prime u_B)$ gives the ratio of electron (including
positron) and magnetic-field energy densities.  Eqs.\ (\ref{Nprime})
and eq.\ (\ref{fes}) imply
\begin{equation}
f_\e^s = {(\delta\e_B)^{(5+p)/2} \over 6\pi d_L^2}\; 
c\sigma_{\rm T} u_{B_{cr}}^2 \tilde k_{eq} V_b^\prime \; {\e_z^{(3-p)/2}
(p-2) \over m_ec^2 \gamma_1^{2-p}}\;,
\label{fes2}
\end{equation}
where $u_{B_{cr}}= B_{cr}^2/8\pi = 7.75\times 10^{25}$ ergs cm$^{-3}$
is the critical magnetic field energy density. The electron injection
index $p$ is related to the observed photon energy index through the
usual relation $\alpha = (p-1)/2$, where $F_\nu \propto
\nu^{-\alpha}$.

Solving for $\delta \e_B = \delta B/B_{cr}$ gives
\begin{equation}
\delta \e_B \equiv y_{eq}= \left [ {9 m_ec^2 d_L^2f^s_\e 
\gamma_1^{2-p}\e_z^{(p-3)/2}\over
	2c \sigma_{\rm T}u_{B_{cr}}^2 (p-2) \tilde k_{eq} r^3_b 
	} \right]^{2/(5+p)}\;,
\label{deb}
\end{equation}
which is a familiar expression from synchro-Compton theory. The
equipartition magnetic field is given when $k_{eq} = 1$, $p = 2$ or
$\alpha = 1/2$, with $(p-2) \gamma_1^{(p-2)}\rightarrow
[\ln(\gamma_2/\gamma_1)]^{-1}= [(1/2)\ln (\e_2/\e_1)]^{-1}$ in 
eq.\ (\ref{deb}), where $\e_1$ and $\e_2$ are the
photon energies bounding the $\alpha = 1/2$ portion of the synchrotron
spectrum.

For a model where the flux at energy $\e$ is assumed to be produced by 
the EC process, one finds from eq.\ (\ref{fet}) that
\begin{equation}
\delta^{3+p} B^2 = {36\pi m_ec^2 d_L^2 f_\e^{EC} \over
	c \sigma_{\rm T}u_* (p-2)\gamma_1^{p-2} \tilde k_{eq} r^3_b } 
\;\left( {2\e_*\over \e_z} \right)^{(3-p)/2}\;.
\label{det}
\end{equation}
For a model where the flux at energy $\e$ is assumed to be produced by 
the SSC process, one finds from eq.\ (\ref{fessc}) that
\begin{equation}
\delta^{(5+p)/2} \e_B^{(9+p)/2} = {27 (m_ec^2)^2 d_L^2 f_\e^{SSC} 
	\gamma_1^{2(2-p)}\e_z^{(p-3)/2}\over
	4c \sigma^2_{\rm T}u_{B_{cr}}^3 (p-2)^2
 \tilde k^2_{eq} r^4_b \Sigma_{\rm C}} \;.
\label{dessc}
\end{equation}

The photon energies $\e$ and $\nu F_\nu$ fluxes (ergs cm$^{-2}$
s$^{-1}$) of knot WK7.8 are ($3.88\times 10^{-11}$, $2.6\times
10^{-15}$) at 4.8 GHz, ($6.96\times 10^{-11}$, $2.95\times 10^{-15}$)
at 8.6 GHz, ($3.48\times 10^{-6}$, $8.6\times 10^{-16}$) at $4.3\times
10^{14}$ Hz, and ($3.1\times 10^{-3}$, $2.5\times 10^{-14}$) at
$3.8\times 10^{17}$ Hz, using the {\it HST} optical values given by
\citet{sch00} and the flux values from Fig.\ 8 of \citet{cha00}. The
redshift $z = 0.651$ for PKS 0637-752, so that $d_L = 1.19\times
10^{28}$ cm for a flat $\Lambda$CDM cosmology with Hubble constant of
72 km s$^{-1}$ Mpc$^{-1}$ and $\Omega_\Lambda = 0.73$, as implied by the WMAP
data \citep{spe03}.  Using these values with a blob size $r_b =
r_{kpc}$ kpc and $p = 2.6$ ($\alpha = 0.8$) implies $\delta B_{\mu
{\rm G}} = 577/( \tilde k_{eq}^{0.26} r_{kpc}^{0.79} \gamma_1^{0.16})$
from eq.\ (\ref{deb}) for the radio synchrotron model, where $B_{\mu
{\rm G}}$ is the comoving magnetic field in $\mu$G. For the X-ray/EC
model, we obtain $\delta =71/[ \tilde k_{eq}^{0.18} r_{kpc}^{0.54}
\gamma_1^{0.11} B_{\mu {\rm G}}^{0.36}]$ using eq.\ (\ref{det}). For the
X-ray SSC model, we obtain $\delta = 2.3\times 10^{5}/[\tilde
k_{eq}^{0.53} r_{22}^{1.05} \gamma_1^{0.32} (\Sigma_{\rm C})^{0.26}
B_{\mu {\rm G}}^{1.53}]$ from eq.\ (\ref{dessc}).

Fig.\ 1 shows the relation between $\delta$ and $B$ for the radio
synchrotron, X-ray EC and X-ray SSC models with $\gamma_1 = 30$ and
$k_{eq} = 1$ for the case of a pair ($k_{pe} = 0$) and e-p ($k_{pe} =
m_p/\gamma_1 m_e \cong 61$) jet.  Increasing $B$ implies an increased
number of electrons to maintain equipartition between the magnetic
field and particle energy. Consequently $\delta$ must decline to
produce the same radio or X-ray fluxes. Because the SSC flux is
proportional to the product of nonthermal electron and magnetic field
energy, $\delta$ declines even faster with $B$ for the X-ray SSC than
for the radio synchrotron model. 

The interceptions of the X-ray EC and X-ray SSC lines with the radio
synchrotron line give solutions satisfying each pair of models. These
solutions also result in {\it about} the minimum total energies
required in these models. The EC model implies values of $B$ of tens
of $\mu$G and requires emission regions in relativistic motion. The
SSC model requires large magnetic fields exceeding mG levels with
debeamed ($\delta \ll 1$) emission.

The $\delta$-$B$ diagram presented here is based on the
equipartition assumption $u_{par}= k_{eq} u_B$ for each process
separately. By contrast, \citet{tav00} assume equipartition for the
synchrotron radio emission, but determine the dependence of $\delta$
and $B$ that reproduces the radio and X-ray data when equipartition is
not assumed. The basic dependences in the two approaches can be
derived and compared in the specific case $p = 3$, corresponding
to a flat $\nu F_\nu$ spectrum.  The bolometric synchrotron luminosity $L_s
\propto u_B u_{par}\delta^4\propto B^4\delta^4$, $u_{par} = u_B\propto 
\delta^4$, so that $B\delta \propto const$. When assuming equipartition
for the EC process, $L_{EC}\propto u_{par}\delta^6\propto
B^2\delta^6$, so that $B\delta^3 \propto const$, as given by eq.\
(\ref{det}). If equipartition is not assumed when jointly satisfying
the synchrotron radio and X-ray EC emission, $L_{EC} \propto
u_{par}\delta^6$, so that $L_{EC}/L_{syn} \propto \delta^2/B^2$,
implying that $\delta\propto B$ (Fig.\ 1 in \citet{tav00}).

The bolometric SSC luminosity $L_{SSC} \propto
u^2_{par}u_B\delta^4\propto B^6\delta^4$ for SSC in equipartition, so
that $B^3\delta^2 \propto const$ when $p = 3$, consistent with eq.\
(\ref{dessc}). If equipartition is not assumed, then $L_s^2/L_{SSC} 
\propto u_B\delta^4 \propto const$ implies $B\delta^2 \propto const$,
which is the dependence that \citet{tav00} obtain. The intersections 
of the lines in both approaches correspond to the $B$ and $\delta$
values where equipartition holds for the radio synchrotron and 
EC/X-ray or SSC/X-ray emission.

\section{Energetics and Minimum Jet Powers}

We use the notations $\fes$, $\fet$ and $\fessc$ for the $\nu F_\nu$
synchrotron, EC, and SSC fluxes measured at photon energies $\e_s$,
$\e_{EC}$ and $\e_{\rm C}$, respectively.  The total comoving energy 
\begin{equation}
W^\prime_{tot} = W^\prime_{par} + W^\prime_B \cong  
{K m_ec^2 (1+k_{pe})\over (p-2)\gamma_1^{p-2}} + 
V_b^\prime u_{B_{cr}}\e_B^2\;
\label{energy}
\end{equation}
is the sum of particle energy, $W^\prime_{par}$, and $B$-field energy,
$W^\prime_B$. The jet power is given by $L_{j} =
\pi r_b^2 \beta\Gamma^2 c W^\prime_{tot}/V^\prime_b$ \citep{cf93}. 
We let $\Gamma \rightarrow \delta$ to derive minimum jet power
$L_{j,min}$.

For a synchrotron model, the jet power 
%\begin{equation}
$L_j(y) = \pi r_b^2 c u_{B_{cr}}[k_{eq}y_{eq}^{(5+p)/2}y^{-(1+p)/2}+y^2]$
%\label{lj}
%\end{equation}
depends only on the product $y\equiv \delta \e_B$ evaluated at radio
energies. The value of $\delta\e_B$ that minimizes the jet power is
given by $\hat y = [(p+1) k_{eq}/4]^{2/(5+p)}y_{eq}$, and $L_{j,min} =
L_j(\hat y)$.  The additional power to produce the X-ray emission
in a two-component synchrotron model is small in comparison with the
radio power, because the X-ray emitting electrons must have large
Lorentz factors and therefore cool very efficiently. The second
electron component, if injected cospatially in a region with the same
magnetic field, must however have a low-energy cutoff so as not to
overproduce optical radiation \citep{ad04}.

For a nonthermal radio synchrotron and an EC X-ray model, jointly
solving eqs.\ (\ref{fes}) and (\ref{fet}) gives
\begin{equation}
\e_B = \delta \left({\fes u_*\over 
\fet u_{B_{cr}}}\right)^{2/(p+1)}\;
\left({\e_{EC}\over 2\e_s\e_*}\right)^{(3-p)/(p+1)}\;,
\label{ebt}
\end{equation}
so that $W^\prime_B \propto \delta^2$.  Using eqs.\ (\ref{Nprime}) and
(\ref{fet}) to solve for $K$ gives
\begin{equation}
W^\prime_{e} =  {m_ec^2\gamma_1^{2-p}
\over \delta^{3+p} (p-2)}\; 
({6\pi d_L^2\fet\over c \sigma_{\rm T}
u_*}) \; [{(1+z)\e_{EC}\over 2\e_*}]^{(p-3) /2},
\label{epart}
\end{equation}
so that $W^\prime_{par} \propto \delta^{-(3+p)}$. The required
particle kinetic energy decreases rapidly with $\delta$ to produce the
observed X-ray flux in the EC model, and the magnetic field energy
increases rapidly to jointly fit the radio synchrotron and EC
X-rays. The Doppler factor $\delta_{EC}$ that minimizes jet power is
given by
$$\delta_{EC}^{5+p} =
{9\over 8}{ (1+k_{pe})(1+p)d_L^2 m_ec^2 (1+z)^{(p-3)/2}\over (p-2)
\gamma_1^{p-2} c \sigma_{\rm T} r_b^3}\;\times$$
\begin{equation}
\left[{(\fet)^{p+5} u_{B_{cr}}^{3-p}\over 
(\fes)^4 u_*^{p+5}}\right]^{1/ (p+1)}\e_s^{{2(3-p)\over p+1 }} 
\left({2\e_*\over \e_{EC}}\right)^{{(3-p)(p+5)\over 2(p+1)}}\;.
\label{deltat}
\end{equation}

We now consider the SSC X-ray model. Jointly solving eqs.\ (\ref{fes})
and (\ref{fessc}) gives
$K  = 2\pi r_b^2
\fessc \; (\e_s/ \e_C)^{(3-p)/2}/(\sigma_{\rm T} \Sigma_{\rm C}\fes )$
and
\begin{equation}
\e_B = \delta^{-{p+5\over p+1}}\;
 \left[ {3d_L^2 \Sigma_{\rm C} \over u_{B_{cr}}c r_b^2}{(\fes)^2\over 
\fessc }\right]^{2\over p+1}\;
\left[{\e_{\rm C}\over (1+z)\e_s^2}\right]^{3-p\over p+1}\;,
\label{ebssc}
\end{equation}from which one obtains
$$
\delta_{SSC}^{2(5+p)/(p+1)} = {8(p-2) 
\gamma_1^{p-2} u_{B_{cr}} r_b \sigma_{\rm T} 
\Sigma_{\rm C}\over 3(p+1)(1+k_{pe}) m_e c^2}\; 
\left({\fes \over \fessc} \right)\;$$
\begin{equation}
\times \left[{3d_L^2  \Sigma_{\rm C}\over u_{B_{cr}} c r_b^2}
{(\fes)^2\over \fessc} \right]^{{4\over p+1}}
\left({\e_{\rm C}\over \e_s}\right)^{{3-p\over 2}}
\left[{\e_{\rm C}\over (1+z)\e_s^2}\right]^{{2(3-p)\over(p+1)}}
\label{deltassc}
\end{equation}
for the Doppler factor $\delta_{SSC}$ that gives the minimum jet power for
the SSC model.

\section{Discussion}

We have derived the minimum jet power $L_{j,min}$ for synchrotron, EC,
and SSC models of the knots and hot spot X-ray emission. The
two-component synchrotron model does not connect the radio and X-ray
fluxes, so that $L_{j,min}$ depends only on the product $\delta
B$. Thus moderate values of $\delta$ are possible. For the parameters
of knot WK7.8 with $\gamma_1 = 30$ and $k_{pe} = m_p/(\gamma_1 m_e)$,
$\hat y =  560(1+k_{pe})^{0.26}/(\gamma_1^{0.16}
r_{kpc}^{0.79})$, implying a minimum jet power of $7\times 10^{46}$
ergs s$^{-1}$.  This is also equal to $L_{j,min}$ for the
synchrotron/EC model, which however only holds for specific values of
$\delta=27$ and $B=36\,\mu$G (see Fig.\ 2). This is because most of
the energy is contained in electrons with $\gamma \sim \gamma_1$,
which is restricted to low values in the EC model. Much larger values
of $\gamma_1$ are allowed in the two-component synchrotron model, so
that even smaller jet powers are possible in this model.

The large value of $\delta = 27$ for the minimum jet power in the
EC model implies small and improbable observing angles $\theta
\leq \delta^{-1}\lesssim 2^\circ$ with a deprojected length of 
$\approx 2$ Mpc of the jet in PKS 0637-752. For larger observing
angles corresponding to $\delta \lesssim 10$, a jet power exceeding
$10^{48}$ ergs s$^{-1}$ is implied (Fig.\ 2).  The jet power could be
reduced assuming a jet composed of e$^+$-e$^-$ plasma. 
The decay of
$\gamma$ rays in the two-component synchrotron model \citep{ad03,ad04}
will produce pairs, but at much higher energies than needed. 
If the
pair plasma were produced in a compact inner jet, then the energy
requirements would be hard to explain because of large adiabatic
losses in the course of expansion of the blob from sub-parsec to kpc
scales (see \citet{cf93} for other arguments against a pair jet).

The SSC model for knot WK7.8 is ruled out. This model formally
satisfies the radio and X-ray fluxes with $\delta_{SSC} = 0.014$
and corresponding magnetic field $B = 0.07$ Gauss when $\Sigma_{\rm C}
=10$, but requires comoving particle and field energies $\gg 10^{61}$
ergs.

The synchrotron/EC model for knot WK7.8 allows a nonvariable X-ray
spectrum that cannot be softer than the radio spectrum, and predicts a
$\gamma$-ray flux at the level of $\approx 0.3\times 10^{-8}$ ph$(>
100$ MeV) cm$^{-2}$ s$^{-1}$. This is detectable at strong
significance with {\it GLAST} in the scanning mode over one year of
observation, but may be difficult to distinguish from the variable
inner jet radiation. The two-component synchrotron model allows
variability at X-ray energies, though at a low level because of the
source size, with nonvarying $\gamma$-ray flux below the {\it GLAST}
sensitivity (see Fig.\ 2 in \citet{ad04}). An interesting study for
{\it GLAST} is to separate a highly variable inner jet component from
a stationary emission component to determine maximum fluxes of the
extended jet.

\vskip0.5in
We thank Dan Schwartz and Andrew Wilson for questions and discussions
about these issues, and the referee for helpful comments on
a different, earlier approach to this problem.  The work of CD
is supported by the Office of Naval Research.  Research of CD and
visits of AA to the NRL High Energy Space Environment Branch are
supported by {\it GLAST} Science Investigation No.\ DPR-S-1563-Y.

\clearpage

%\vskip-6.0in
\begin{figure}[t]
%\includegraphics[width=4.0 cm]{fig1.eps}
%\vskip-4.0in
\plotone{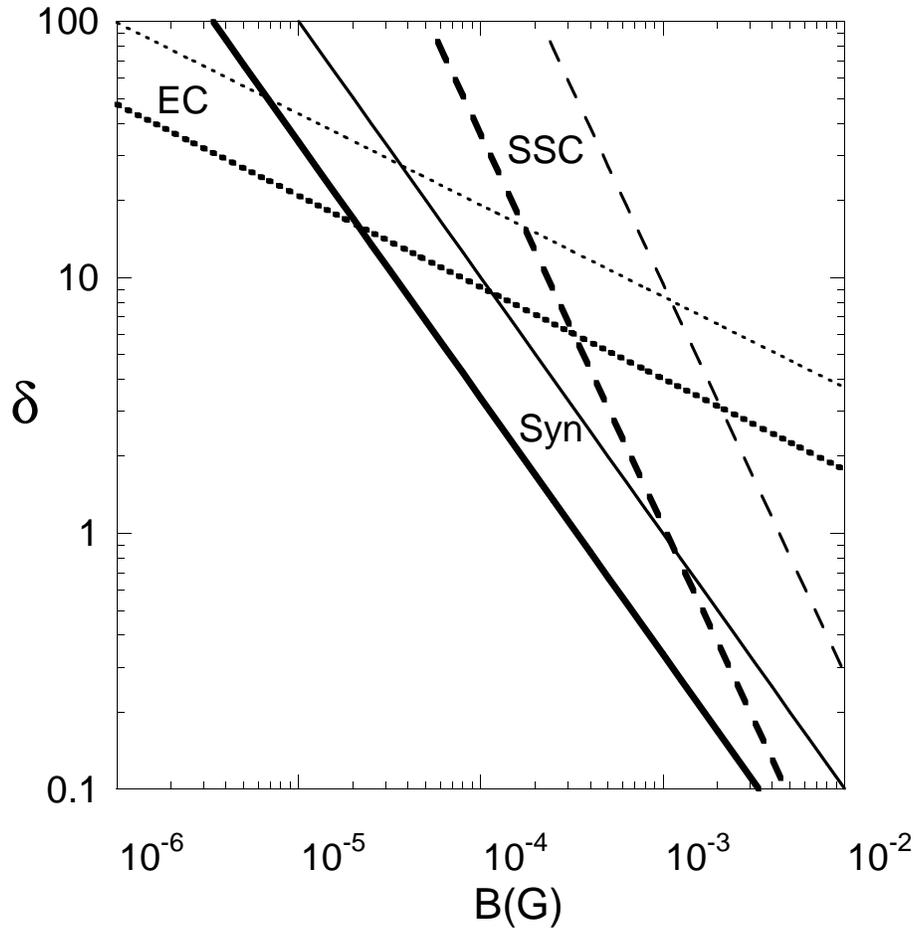}
\vskip-1.5in
\caption{Dependence of Doppler factor on magnetic field to produce 
radio synchrotron flux (solid lines), X-ray flux from
Compton-scattered CMBR (dotted lines), and X-ray flux from the SSC
process (dashed lines) observed from knot WK7.8 under the
condition of equipartition between magnetic field and total particle
energy densities ($k_{eq} = 1$). The minimum electron Lorentz
factor $\gamma_1 = 30$, $\Sigma_{\rm C} = 10$, $r_b = 1$ kpc, 
and heavy and light curves are for a pair and 
e-p plasma, respectively. }
\end{figure}
\clearpage

%\vskip-6.0in
\begin{figure}[t]
%\includegraphics[width=4.0 cm]{fig2.eps}
%\vskip-4.0in
\plotone{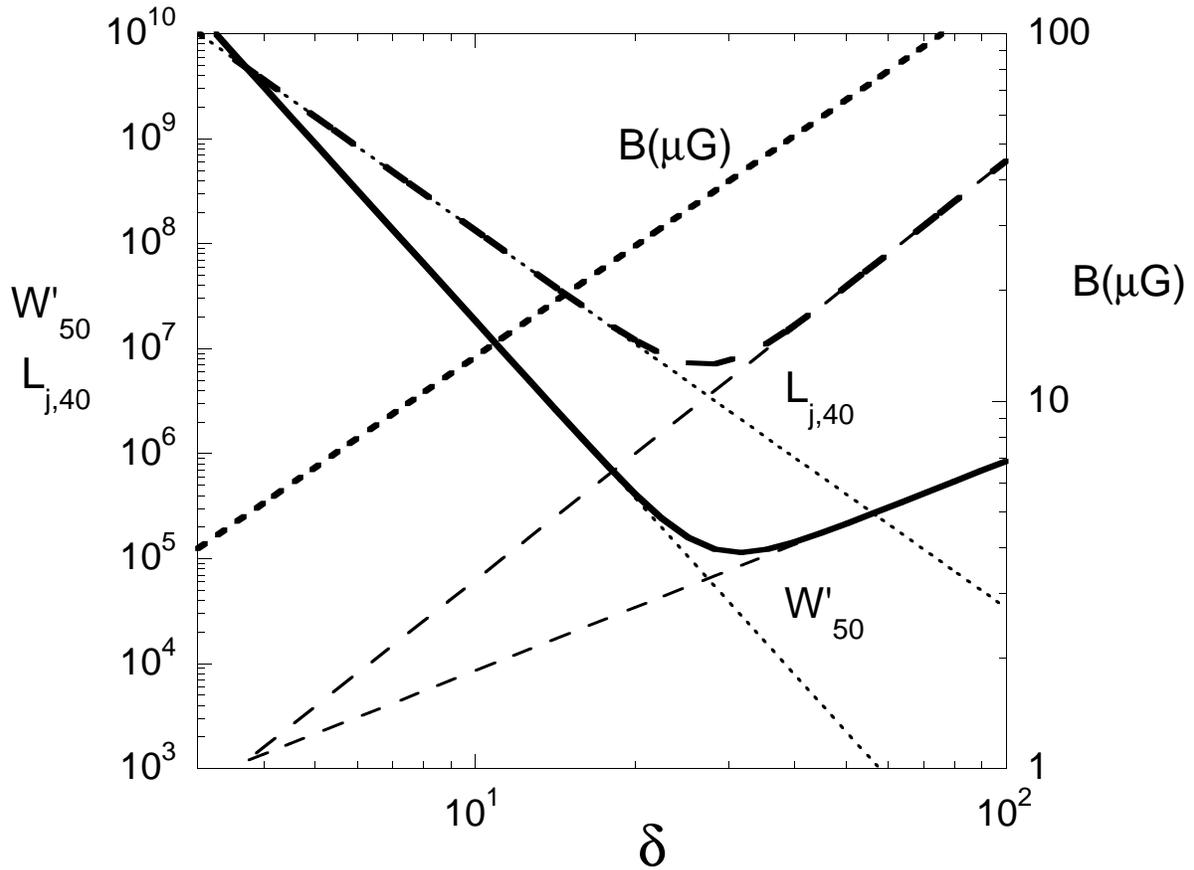}
\vskip-1.5in
\caption{Total comoving energy $W^\prime_{50}$ in units 
of $10^{50}$ ergs (solid) and jet power $L_{j,40}$ in units of
$10^{40}$ ergs s$^{-1}$ (long-dashed) as a function of $\delta$,
divided into particle (dotted) and magnetic field (dashed) components
for a radio synchrotron and X-ray EC model of knot WK7.8 of PKS
0637-752. The comoving magnetic field is shown by the heavy dotted
line. The minimum electron Lorentz factor $\gamma_1 = 30$,
$\Sigma_{\rm C} = 10$, $r_b = 1$ kpc, and we consider a jet made of
e-p plasma. }
\end{figure}

\end{document}